\documentclass[10pt,letterpaper,twocolumn]{article} 
\usepackage{ol2} 
\usepackage[english]{babel}                 
\usepackage{amsmath}                        
\usepackage{setspace}

\newcommand{\ket}[1]{\left| #1\right\rangle}

\newcommand{\ketbra}[2]{\left| #1\right\rangle\!\left\langle#2\right|}

\newcommand{\iea}[0]{{\it et al.~}}
\newcommand{\ieac}[0]{{\it et al., }}
\newcommand{\eeqref}[1]{Eq.~(\ref{#1})}

\definecolor{MyGreen}{rgb}{0.0,0.5,0.1}

\begin{document}

\twocolumn[

\title{Instant single-photon Fock state tomography}


\author{S. R. Huisman,$^{1,2}$ Nitin Jain,$^1$ S. A. Babichev,$^{1,3}$ Frank Vewinger,$^{1,4}$, A. N. Zhang,$^{1,5}$ S. H. Youn,$^{1,6}$ and A. I. Lvovsky$^{1,*}$}
\address{$^1$Institute for Quantum Information Science, University of Calgary, Calgary, Alberta T2N 1N4, Canada\\
$^2$Complex Photonic Systems (COPS), Faculty of Science and Technology and MESA+ Institute for Nanotechnology,
University of Twente, 7500 AE Enschede, The Netherlands\\
$^3$Power Wind GmbH, 20537 Hamburg, Germany\\
$^4$Institut f\"{u}r Angewandte Physik, Universit\"at Bonn, 53115 Bonn, Germany\\
$^5$Department of Physics, East China Normal University, Shanghai 200062, China\\
$^6$Department of Physics, Chonnam National University, Gwangju, 500-757, Korea\\
Corresponding author: lvov@ucalgary.ca}

\maketitle

\begin{abstract}Heralded single photons are prepared at a rate of $\sim 100$ kHz via conditional measurements on polarization-nondegenerate biphotons produced in a periodically poled KTP crystal. The single-photon Fock state is characterized using high frequency pulsed optical homodyne tomography with a fidelity of $(57.6 \pm 0.1)\%$. The state preparation and detection rates allowed us to perform on-the-fly alignment of the apparatus based on real-time analysis of the quadrature measurement statistics. \end{abstract}
]


The ability to produce single-photon states is of importance for optical quantum computation \cite{KLM}, quantum cryptography \cite{qcrypt-review}, and quantum state engineering \cite{engineering}. Many of these applications require the photons to be generated with a high fidelity in a pure, well-defined spatiotemporal mode. This requirement can be approached by heralded single photons prepared via conditional measurements on biphotons produced due to spontaneous parametric down-conversion (SPDC) \cite{BurnHamWeinberg, HongMandel1986}. In this method, photon pairs produced in a nondegenerate optical parametric amplifier are split into two channels: \textit{trigger} and \textit{signal}. A single spatiotemporal mode is selected in the trigger channel and subjected to measurement with a single-photon counting module (SPCM). A detection event heralds preparation of the single photon in the conjugate spatiotemporal mode of the signal channel \cite{Lvovsky2001, Aichele2002}.

A significant disadvantage of heralded photons is that they are not prepared on demand. The bimodal state of light produced due to each pump pulse (we specialize to the pulsed case) obeys thermal statistics:
\begin{equation}\label{NOPA}
\ket\Psi\approx(1-\gamma^2)\left(\ket{0,0}+\gamma\ket{1,1}+\gamma^2\ket{2,2}+\ldots\right),
\end{equation}
where the numbers refer to Fock states in the signal and trigger channels and $\gamma$ is proportional to the pump field amplitude as well as the nonlinear susceptibility of the down-conversion medium. In the weak pump regime ($\gamma\ll 1$), a trigger photon detection event indicates that the pulse in the signal channel contains a single photon. The probability of this event scales as $\propto\gamma^2$. If multiple photons need to be generated from $n$ heralded sources simultaneously, the probability scales as $\propto\gamma^{2n}$, which results in unpractically long data acquisition times. A higher pump intensity increases the biphoton production rate, but leads, with non-ideal photon detectors, to a spurious multiphoton component in the signal, and thus to reduced fidelities. A compromise between these two regimes is facilitated by higher pump pulse repetition frequencies. In this way, a reasonably high photon creation rate can be reached while maintaining a sufficiently low ratio between $(n+1)$- and $n$- photon events. More frequent pump pulses however imply a reduced optical energy per pulse, which can be compensated by using novel materials with high nonlinear susceptibilities. Additionally, the quantum state measurement optoelectronics must be sufficiently fast to distinguish between neighboring laser pulses.

In this Letter, we report homodyne tomography \cite{LvovskyRaymer} of the single-photon Fock state $\ket{1}$ generated under such intermediate conditions. The preparation rate is on the scale of 100 kHz with a pump pulse frequency of 76 MHz. This means, on one hand, that the two- and three-photon preparation rates are on the scale of 100 and 0.1 Hz, respectively, which is not prohibitively low. On the other hand, the fraction of multiphoton events heralded as single photons is on the order of 1\% (assuming a 10\% trigger photon detection efficiency), which is acceptable for most applications. A particularly attractive feature of our experiment is that the amount of data sufficient for reliable state reconstruction is acquired within a fraction of a second, permitting on-the-fly optimization of the setup.

Our setup can be compared with that described by Ourjoumtsev \iea \cite{Ourjoumtsev2006}. In that experiment, a cavity dumped Ti:Sapphire laser was employed to increase the pump intensity at the cost of decreasing the repetition frequency of the pump pulses to 800 kHz. Preparation rates of $\sim 4500$ Hz and $\sim 15$ Hz were reported for Fock states $\ket 1$ and $\ket 2$, respectively.

\begin{figure}[tbp]
\centering
\includegraphics[width=0.9\columnwidth]{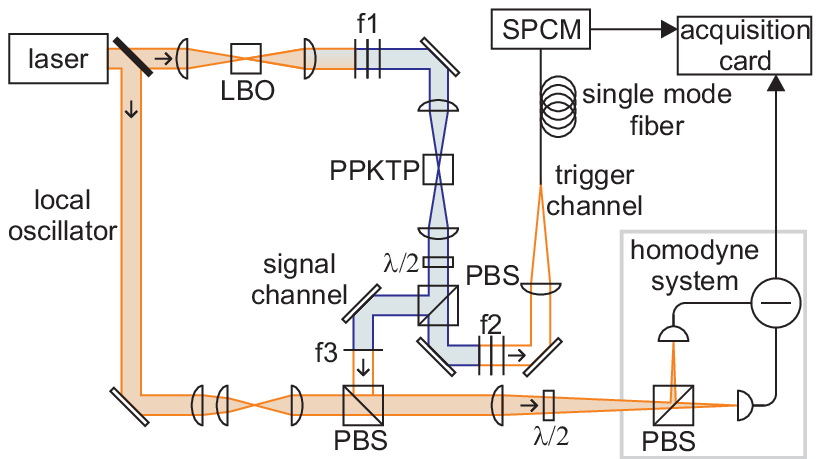}
\caption{\small (Color online) Experimental setup. f1: Spatial and spectral filters transmiting the pump light (blue) with $\lambda \sim 395$ nm. f2: Spectral filters transmitting states with $\lambda \sim 790$ nm and $\Delta \lambda = 0.3$ nm. f3: lowpass filter blocking the pump light (Semrock BLP01-635R-25).}
\label{fig:set}
\end{figure}

Our apparatus is based on Ref.~\cite{Lvovsky2001} and displayed in Fig. \ref{fig:set}. A mode-locked Ti:Sapphire laser (Coherent Mira 900) emitted transform-limited pulses at $\sim 790$ nm with a repetition rate of 76 MHz and a pulse width of $\sim 1.7$ ps. Most of the light was frequency doubled in a single pass through a 17 mm long LBO crystal, yielding typically $\sim 75$ mW average second harmonic power after spatiotemporal filtering (f1). This field was coupled into a 2-mm long periodically poled potassium-titanyl phosphate (PPKTP) crystal to pump type-II collinear SPDC. The lengths of the crystals were chosen so that the group velocity dispersion would not lead to significant changes of the pulse width in both nonlinear transformations.

The generated photon pair was separated into the signal and trigger channels according to the photons' polarizations using a $\lambda/2$ plate and a polarizing beam splitter (PBS). A spatiotemporal mode for the trigger photons was selected with a 0.3-nm wide spectral filter (f2) centered at the laser wavelength followed by a single mode fiber. The signal state was prepared in the conjugate mode when a SPCM (Perkin Elmer, SPCM-AQR-14-FC) registered the trigger photons. By temporarily blocking the pump field, we estimated that $\sim 2 \%$ of the trigger events occured due to scattered laser light, corresponding to a preparation efficiency of $\eta_p \sim 0.98$. Trigger event rates up to 140 kHz were routinely observed.

A fraction of the original laser beam was split off before frequency doubling to provide the local oscillator (LO) for balanced homodyne detection. The local oscillator phase was allowed to vary randomly. The homodyne detector was aligned by coupling a 790 nm alignment beam into the trigger channel of parametric down-conversion. Classical difference frequency signal between the pump and alignment fields simulated the optical mode of the prepared signal photon \cite{Aichele2002, Klyshko}, and was matched to the LO by optimizing the interference visibility therewith. By implementing a zoom lens configuration in the LO path, a visibility of $\sim 85 \%$ was reached, corresponding to a mode-matching efficiency of $\eta_m \sim 73 \%$ \cite{Aichele2002}. The linear losses in the signal channel were mainly due to the lowpass filter (f3 in Fig.~1) and corresponded to an efficiency of $\eta_l \sim 96\%$.

The homodyne detector circuit employed two Si-Pin photodiodes (Hamamatsu S5972) in a low-noise electronic substraction circuit. These photodiodes possessed a quantum efficiency $\eta_D = 85 \%$, which is slightly lower than photodiodes typically used in high-speed homodyne detector circuitry in the past \cite{Hansen2001,LvovskyRaymer,Zavatta2002, Zavatta 2004, Zavatta2006}. However, their lower capacitance and dark current permitted us to significantly improve the bandwidth and signal-to-noise characteristics. The difference photocurrent was processed by an operational amplifier (OPA 847) in the transimpedance gain configuration. The detector had a $-3$ dB bandwidth at $\sim 90$ MHz, exceeding the repetition frequency of the LO pulses. The shot noise, measured at an average LO power of 25 mW, was $\sim 14$ dB higher than the electronic noise, corresponding to an electronic quantum efficiency of $\eta_{el} \sim 93 \%$ \cite{Appel}.

The amplified difference photocurrent signal produced with the homodyne detector was digitized with an acquisition card (Agilent Acqiris DP211), integrated and displayed in real time. With each trigger event, 128-ns data segments were acquired. In this way, quadrature measurements were collected for the signal pulse as well as 8 neighboring pulses, whose state was approximately vacuum, in order to calibrate the detector \cite{Appel}. The acquisition rate, about 25,000 segments per second, was limited by the information transfer rate between the acquisition card and the computer. The acquired quadrature data corresponds to the state $\hat\rho  = (1-\eta)\ketbra 0 0  + \eta \ketbra 1 1$, where $\eta$ is the overall quantum efficiency \cite{Lvovsky2001}.

The variance of the acquired quadratures is then given by $\langle Q^2\rangle=\frac{1}{2}+\eta$ (using a scaling convention in which $\langle Q^2\rangle=\frac{1}{2}$ for the vacuum state). The value of $\eta$ obtained from the variance was calculated and displayed periodically on acquisition of a specific number of data segments, typically 4000 to 6000 (corresponding to a display update rate of 4 to 5 times a second). The alignment of various parts of the setup could be improved using the displayed efficiency as a guide. Typically, starting from a preliminary alignment using classical fields, we were able to increase the efficiency by 5--10\%.




\begin{figure}[tbp]
\centering
\includegraphics[width=1\columnwidth]{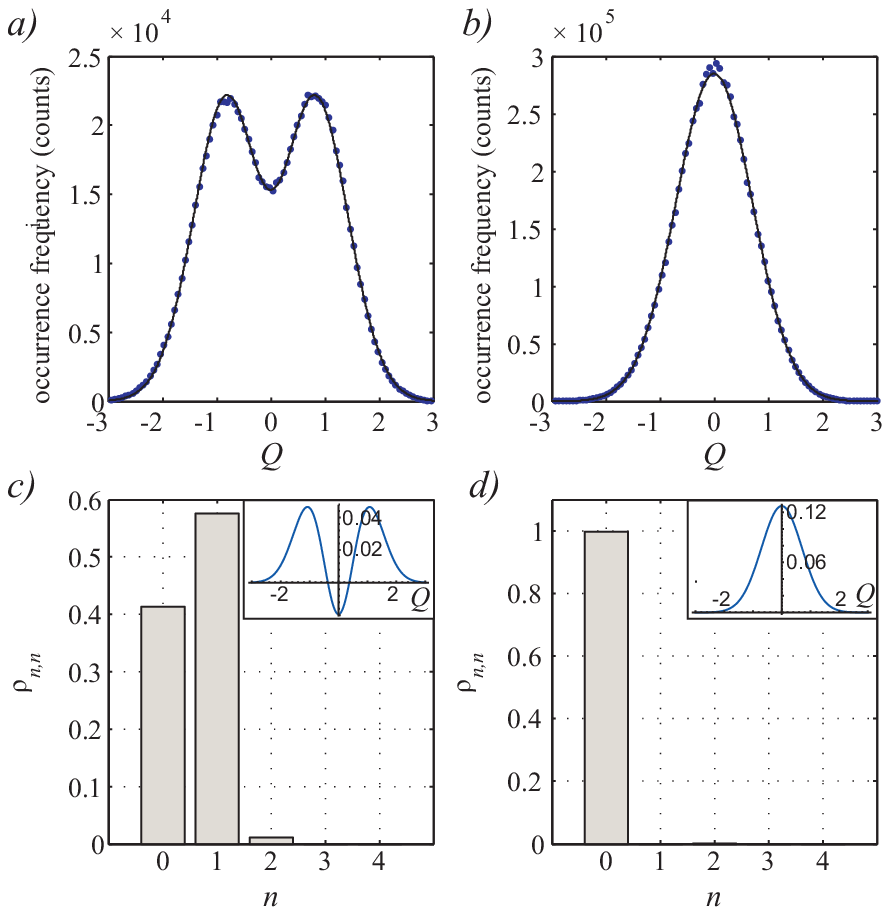}
\caption{\small (Color online) Experimental results. Top row: histogram of the collected signal state (a) and vacuum state (b) quadratures. The black lines are scaled marginal distributions based on the reconstructed density matrices. Bottom row: Reconstructed diagonal density matrix elements for the signal (c) and vacuum (d) states. The insets show cross-sections of the corresponding Wigner functions.}
\label{fig:quadratures}
\end{figure}

Upon optimization of the setup, a large data set of $10^6$ points was acquired and subjected to detailed analysis [Fig.~\ref{fig:quadratures}(a,b)]. First, we corrected for residual correlations caused by a finite bandwidth of the homodyne detector, which lead to an additional quantum efficiency improvement of 0.5\% on average. The maximum likelihood method \cite{Banaszek98,LvovskyRaymer,Lvovsky2004} was then used to reconstruct the diagonal density matrix elements shown in the left column of Table 1 and in Fig.~\ref{fig:quadratures}(c). Contributions of higher number states were negligible.

The vacuum state data yielded the density matrix elements displayed in the right column of Table 1 with negligible errors [Fig.~\ref{fig:quadratures}(d)].
The presence of the two-photon component in the vacuum state estimation is due to the persistent non-Gaussian ``hump'' around $Q=0$ in the marginal distribution [Fig.~\ref{fig:quadratures}(b)] which we believe to arise due to a finite digitizer resolution.

\begin{table}[t]
\centering
\caption{\small Quantum state reconstruction results. The uncertainties shown are standard deviations of the state estimation based on the Fisher information matrix \cite{HradilNJP}.}
\begin{tabular}{|c|c|c|}
\hline
 & Fock & vacuum \\
\hline
$\rho_{00}$
 & $0.4138 \pm 0.0006$ & 0.9987 \\
$\rho_{11}$
 & $0.5758 \pm 0.0011$ & 0 \\
$\rho_{22}$
 & $0.0104 \pm 0.0006$ & 0.0013 \\
\hline
\end{tabular}
\end{table}

The noticeable value for $\rho_{2,2} = 0.0104 \pm 0.0006$ in the Fock state density matrix is likely to be caused by multipair creation events in parametric down-conversion. In order to evaluate the suitability of our setup for multiphoton quantum technology experiments, let us estimate $\gamma$. Assuming the quantum efficiency of the SPCM, including linear losses in the trigger channel, to equal $\eta_t\ll 1$, the state heralded by the photon detection event, neglecting dark counts, is given, according to \eeqref{NOPA}, by $\hat\rho_h \propto \eta_t\gamma^2\ketbra 1 1 + 2\eta_t\gamma^4\ketbra 2 2 + O(\gamma^6)$. If this state is subjected to an optical loss $1-\eta$, it becomes $\hat\rho \propto \eta_t[(1-\eta)\gamma^2+2(1-\eta)^2\gamma^4]\ketbra 00 + \eta_t[\eta\gamma^2+4\eta(1-\eta)\gamma^4]\ketbra 11+2\eta_t\eta^2\gamma^4\ketbra 2 2 + O(\gamma^6)$. Normalizing and comparing this result with the density matrix obtained, we find $\eta=0.5787\pm0.0011$ and $\gamma^2\approx0.0160\pm0.0010$.

The probability of the heralding event is  $p_t\approx\eta_t\gamma^2$. In this particular data set, the trigger event frequency was 88 kHz, which corresponds to $p_t=1.15 \cdot 10^{-3}$, from which we find $\eta_t\approx 7$\%, which is consistent with direct estimate of the trigger channel transmission.


In summary, we have reported homodyne reconstruction of the single-photon Fock state produced using conditional measurements on photon pairs generated via type-II spontanteous parametric down-conversion in a periodically poled KTP crystal with a detection rate on the scale of 100 kHz. The single-photon reconstruction fidelity is comparable to that reported previously \cite{Lvovsky2001,Ourjoumtsev2006,Zavatta2004}. The combination of the preparation rate as well as fast measurement and analysis permitted virtually instantaneous tomographic characterization, so optimization of the experimental setup in real time was possible. The achieved parameters of the experimental setup permit implementation of various new experiments in quantum-optical engineering and information processing that require simultaneous operation of two and three single-photon sources.

This work was supported by Natural Sciences and Engineering Research Council of Canada, Canadian Institute for Advanced Research, Alberta Ingenuity Fund, Canada Foundation for Innovation, and Quantum{\it Works}. We thank M. Lobino for helpful discussions.


\begin{thebibliography}{10}
\newcommand{\enquote}[1]{``#1''}

\bibitem{KLM} E. Knill, R. Laflamme, G. J. Milburn, Nature {\bf 409}, 46 (2001).

\bibitem{qcrypt-review} N. Gisin, G. Ribordy, W. Tittel, and H. Zbinden,  Rev. Mod. Phys. {\bf 74}, 145 (2002)

\bibitem{engineering} F. Dell'Anno, S. De Siena, and F. Illuminati, Phys. Rep.
{\bf 428}, 53 (2006).

\bibitem{BurnHamWeinberg}
D. C. Burnham and D. L. Weinberg,
Phys. Rev. Lett. \textbf{25}, 84 (1970).

\bibitem{HongMandel1986}
C. K. Hong and L. Mandel,
Phys. Rev. Lett. \textbf{56}, 58 (1986).

\bibitem{Lvovsky2001}
A. I. Lvovsky, H. Hansen, T. Aichele, O. Benson, J. Mlynek and S. Schiller,
Phys. Rev. Lett. \textbf{87}, 050402 (2001).

\bibitem{Aichele2002}
T. Aichele, A.I. Lvovsky and S. Schiller,
Eur. Phys. J. D {\bf 18}, 237–245 (2002).

\bibitem{LvovskyRaymer}
A. I. Lvovsky and M. G. Raymer
Rev. Mod. Phys. {\bf 81}, 299 (2009)

\bibitem{Ourjoumtsev2006}
A. Ourjoumtsev, R. Tualle-Brouri and P. Grangier,
Phys. Rev. Lett. \textbf{96}, 213601 (2006).

\bibitem{Klyshko}
D. N. Klyshko, Phys. Lett. A \textbf{132}, 299 (1988).



\bibitem{Hansen2001}
H. Hansen, T. Aichele, C. Hettich, P. Lodahl, A. I. Lvovsky, J. Mlynek and S. Schiller,
Opt. Lett.  \textbf{26}, 1714 (2001).


\bibitem{Zavatta2002}
A. Zavatta, M. Bellini, P. L. Ramazza, F. Marin, and F. T.
Arecchi, J. Opt. Soc. Am. B {\bf 19}, 1189 (2002).

\bibitem{Zavatta2004}
A. Zavatta, S. Viciani and M. Bellini,
Phys. Rev. A \textbf{70}, 053821 (2004).

\bibitem{Zavatta2006}
A. Zavatta, S. Viciani and M. Bellini,
Laser Phys. Lett. \textbf{3}, 3 (2006).


\bibitem{Appel}
J. Appel, D. Hoffman, E. Figueroa and A. I. Lvovsky
Phys. Rev. A \textbf{75}, 035802 (2007).

\bibitem{Banaszek98} K. Banaszek, Phys. Rev. A 57, 5013 (1998).

\bibitem{Lvovsky2004}
A.I. Lvovsky,
J. Opt. B: Q. Semiclass. Opt. \textbf{6} S556 (2004).

\bibitem{HradilNJP} J. \v{R}eh\'{a}\v{c}ek \ieac New J. Phys. {\bf 10}, 043022 (2008).

\end{thebibliography}
\end{document}